\documentclass[twocolumn]{aastex631}

\begin{document}
\title{SMA and NOEMA reveal asymmetric sub-structure in the protoplanetary disk of IRAS~23077+6707}

\author[0000-0002-4248-5443]{Joshua B. Lovell}
\affiliation{Center for Astrophysics, Harvard \& Smithsonian, 60 Garden Street, Cambridge, MA 02138-1516, USA}

\author[0000-0002-8623-9703]{Leon Trapman}
\affiliation{Department of Astronomy, University of Wisconsin-Madison, 
475 N Charter St, Madison, WI 53706, USA}

\author[0000-0002-5688-6790]{Kristina Monsch}
\affiliation{Center for Astrophysics, Harvard \& Smithsonian, 60 Garden Street, Cambridge, MA 02138-1516, USA}

\author[0000-0003-2253-2270]{Sean M. Andrews}
\affiliation{Center for Astrophysics, Harvard \& Smithsonian, 60 Garden Street, Cambridge, MA 02138-1516, USA}

\author[0000-0003-2014-2121]{Alice S. Booth}
\affiliation{Center for Astrophysics, Harvard \& Smithsonian, 60 Garden Street, Cambridge, MA 02138-1516, USA}

\author[0000-0002-3490-146X]{Garrett K. Keating}
\affiliation{Center for Astrophysics, Harvard \& Smithsonian, 60 Garden Street, Cambridge, MA 02138-1516, USA}

\author[0000-0003-4902-222X]{Takahiro Ueda}
\affiliation{Center for Astrophysics, Harvard \& Smithsonian, 60 Garden Street, Cambridge, MA 02138-1516, USA}

\author[0000-0003-1526-7587]{David J. Wilner}
\affiliation{Center for Astrophysics, Harvard \& Smithsonian, 60 Garden Street, Cambridge, MA 02138-1516, USA}


\begin{abstract}
We present high--resolution data of IRAS~23077+6707 (`\textit{Dracula's Chivito}') with the Submillimeter Array (SMA at 1.33\,mm/225.5\,GHz) and the Northern Extended Millimeter Array (NOEMA at 2.7\,mm/111.7\,GHz and 3.1\,mm/96.2\,GHz).
IRAS~23077+6707 is a highly-inclined and newly discovered protoplanetary disk, first reported in 2024.
We combine SMA baselines from the Compact, Extended and Very Extended arrays, and NOEMA baselines from its A and C configurations, and present continuum images with resolution ${\lesssim}0.8''$, which constitute the first sub--arcsecond resolution maps of IRAS~23077+6707.
The images show extended linear emission that spans $5.6{-}6.1''$ as expected for a radially extended, highly-inclined protoplanetary disk.
Accompanied with lower resolution data, we show that the disk has a steep spectral index, ranging from $\alpha=3.2{-}3.9$.
We present evidence of multiple radial emission peaks and troughs in emission, which may originate in disk rings and a central cavity.
We further present evidence that these radial structures are asymmetric; hosting a a significant brightness asymmetry, with emission enhanced by up to 50\% in the north versus the south.
We discuss hypotheses about the potential origins of these features, including the possibility that IRAS~23077+6707 hosts a rare example of an eccentric protoplanetary disk, which can induce these radially asymmetric structures.
We present a simple eccentric continuum model of IRAS~23077+6707, and show for an eccentricity of $e \approx 0.26$, that this can reproduce the bulk morphology of the emission.
\end{abstract}


\section{Introduction} \label{sec:intro}
Protoplanetary disks are dust- and gas-rich circumstellar disks present around young stars, and are expected to be the main sites in which planets form \citep{WilliamsCieza11, Andrews20, Drazkowska22}.
Recent advances in millimeter interferometry, particularly with the Atacama Large Millimeter/submillimeter Array (ALMA), have brought into clear view the continuum structures of these disks, which are now known to host radial sub-structures such as rings, gaps and cavities \citep[see e.g.,][]{ALMAPart15, Andrews+16, Andrews18, Long19}.
More rarely, asymmetric sub-structures such as clumps and vortices (collectively known as arcs), spirals, and eccentricities have been found \citep[see above, and examples such as][]{vdMarel+13, Benisty+15, Perez+16, Huang+18b, Benisty+21, Ragusa+2021, vdMarel+21, Yang+2023}.
Many mechanisms have been identified to explain such symmetric and asymmetric sub-structures \citep[including (magneto)-hydrodynamical, dynamical, gravitational, and chemical, see e.g.,][and references therein]{Andrews20} hence their study is shedding light on how disks evolve during the epoch of planet formation, and how planets may be shaping such disks. 
Indeed in a handful of protoplanetary disks, direct detections of protoplanets and/or circumplanetary disks have now been confirmed \citep[see e.g.,][]{Isella+2019, Benisty+21, Hammond+2023}.

\begin{figure*}
    \centering
    \includegraphics[width=1.0\linewidth]{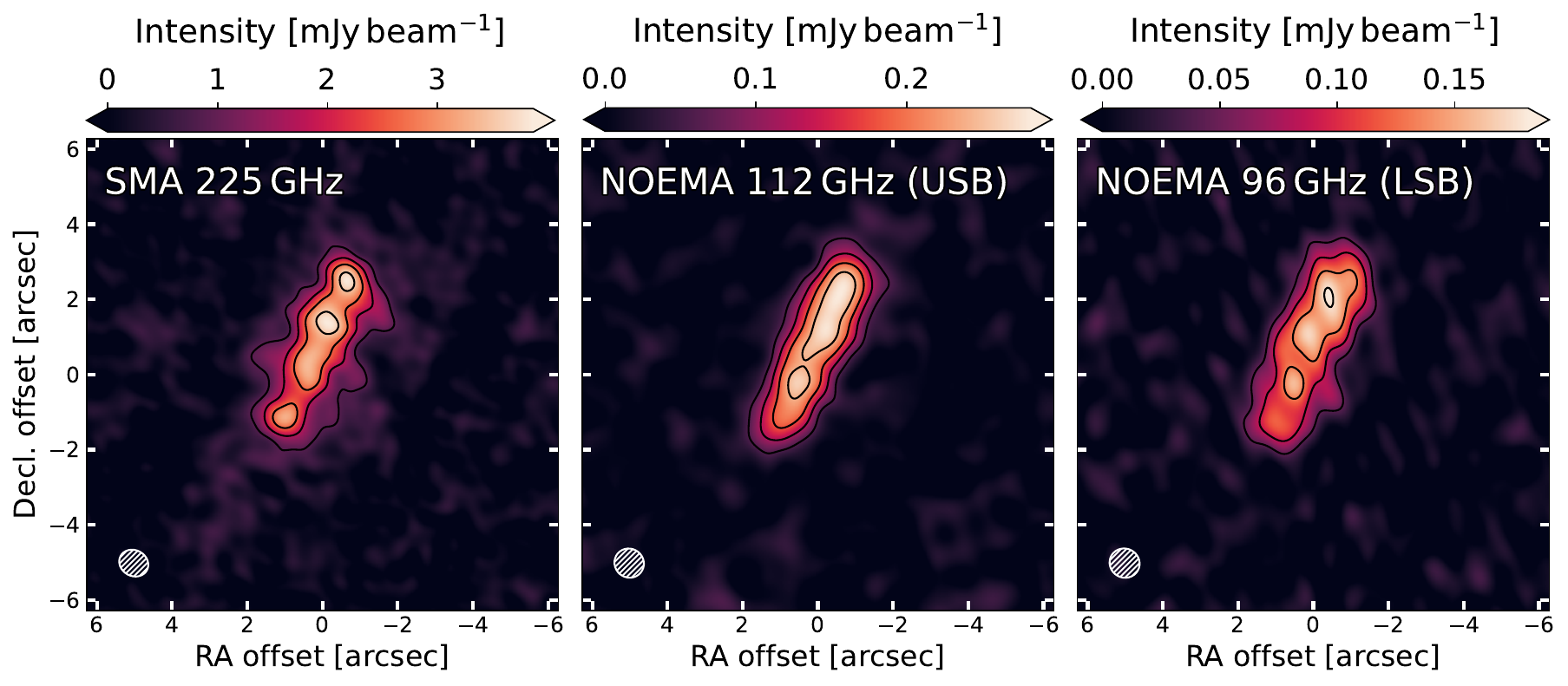}
    \caption{SMA (left), NOEMA upper-sideband (middle) and NOEMA lower-sideband (right). We present SMA and NOEMA contours at of 4, 8, 12, and 16$\sigma$ levels. Relative offsets are with reference to the phase center of the observations. Synthesised beams are presented in the lower-left of each panel, with sizes of $0\farcs80{\times}0\farcs69$, $0\farcs80{\times}0\farcs76$ and $0\farcs81{\times}0\farcs77$ and position angles of $61^\circ$, $50^\circ$, and $60^\circ$ respectively. }
    \label{fig:main}
\end{figure*}

The recent discovery of the giant (${>}1000\,$au extent) edge-on disk of IRAS~23077+6707 \citep{Berghea24, Monsch24} presented a number of questions that only higher-resolution follow-up observations could address.
In particular, how does this extended disk compare to the broader population of protoplanetary disks? 
Does the millimeter continuum present any sub-structures? 
And are asymmetries, noted by \citet{Monsch24} in the scattered light images, present in the thermal emission? 
Whilst ALMA has successfully uncovered continuum sub-structures in disks typically in the southern hemisphere, the opportunity to resolve sub-structures in extended disks (with major axes ${\gtrsim}1''$) in the northern hemisphere is enabled by both the Submillimeter Array (SMA) and the NOrthern Extended Millimetre Array (NOEMA).

We present the first results of a high-resolution millimeter follow-up campaign to resolve the disk surrounding IRAS~23077+6707 with the SMA and NOEMA.
We present in \S\ref{sec:data} the SMA and NOEMA observations including their data reduction calibration routines. 
In \S\ref{sec:analysis} we present the analyses of these data, and in \S\ref{sec:discussion} we discuss IRAS~23077+6707 in the context of other recent studies of edge-on protoplanetary disks, and consider origins for the asymmetric radial sub-structure.
We summarise our findings in the conclusions section \S\ref{sec:conclusions}.

\section{Observations of IRAS~23077+6707} \label{sec:data}
\subsection{SMA observations \& data calibration}
The Submillimeter Array (SMA) is an 8-antenna (sub-)millimeter interferometer based on Maunakea, Hawai'i \citep{Ho2004}.
We present observations of IRAS~23077+6707 from three dates in setups as outlined in Table~\ref{tab:obs} in projects 2022B-S054 (PI: K. Monsch), 2023A-S052 (PI: J. B. Lovell) and 2024A-S040 (PI: J. B. Lovell).
In total, IRAS~23077+6707 was observed for 18.3\,hrs (on-source), with the SMA's `Compact' (COM), `Extended' (EXT) and `Very Extended' (VEX) configurations.
All observations had SMA pointing center coordinates (J2000) of 23:09:43.645 (RA) and +67:23:38.940 (Decl.). 

The observations were conducted with the SMA `Wideband Astronomical ROACH2 Machine' (SWARM) correlator, comprising two independent receivers with two $12\,$GHz sidebands, each with 140\,kHz channel resolution  \citep[see][]{Primiani+2016J}.
In Table~\ref{tab:obs} we specify the observation setup per-track including details of calibrators. 
In all three setups, both receivers were tuned to central local oscillator (LO) frequencies of 225.538\,GHz ($\lambda=1.33\,\mathrm{mm}$).
We converted the raw SMA data to the \textit{Common Astronomy Software Applications} ({\tt CASA}) measurement set format \citep{McMullin+2007}, re--binning channels by 256 to reduce file sizes for this continuum study with {\tt pyuvdata} version 3.1.3 \citep{Hazelton+2017, pyuvdata_Karto2025}.
Auto-flagging and gain-phase, gain-amplitude, bandpass, and flux solution tables were calculated with the SMA `\textit{COMPASS}' (Calibrator Observations for Measuring the Performance of Array Sensitivity and Stability) tool, version 0.11.0 (Keating et al., in prep.)\footnote{Originally designed as an engineering tool that leverages calibrator data to determine the health of the SMA systems, \textit{COMPASS} has been purposefully grown into a science-focused data reduction pipeline, with the goal of delivering users ``ready-for-imaging'' data sets.}.
Before each calibration, we manually checked for any remaining interference spurs using {\tt CASA plotms} and flagged a small time range of BL~Lac channels in time which appeared spurious\footnote{We flagged BL~Lac data over the UTC range 04:14:00 to 04:20:00 (2023/10/07), which in comparison to data from the UTC range 04:21:00 to 04:53:00 was lower in amplitude by ${\sim}10\%$. This procedure retained 64/69 bandpass calibrator scans, sufficient to accurately calibrate the spectral/frequency response of the science target data.}. 
Overall this demonstrated the success of \textit{COMPASS}'s automated flagging solution.
We additionally flagged the channels containing bright $^{12}$CO and $^{13}$CO $J{=}2{-}1$ line emission to avoid any gas emission influencing analysis of the continuum (dust) emission, which are known to have integrated line fluxes of ${\approx35}$\,Jy\,km\,s$^{-1}$ and ${\approx14}$\,Jy\,km\,s$^{-1}$ respectively \citep[see][]{Monsch24}. 
We note that neither of these lines were visible by eye in the {\tt CASA plotms} view of the corrected (pre-flagged) data for IRAS~23077+6707, and thus likely contributed negligible flux to the continuum.
We used the SMA standard reduction script in {\tt CASA} version 6.6.3\footnote{The SMA {\tt CASA} reduction script can be accessed via: \url{https://github.com/Smithsonian/sma-data-reduction}.} to apply calibration tables to each measurement set independently.
Phases and amplitudes (as functions of time and frequency) for all post-calibration science and calibrator were inspected, which showed exceptional stability and band-to-band consistency.

Calibrated ({\tt CASA} `corrected') data for IRAS~23077+6707 were extracted for each measurement set separately using the {\tt CASA mstransform} tool, with additional channel binning to 4 channels per spectral window, minimizing bandwidth smearing whilst also minimizing data volumes\footnote{No time binning was applied after we found a critical error in how this re-scaled the interferometric weights. We report this here as a caution to others.}.
One can assess the extent of interferometric bandwidth smearing by utilizing equations~18-17 and 18-24 from \citet{Bridle+1999}, which we provide here in the form
\begin{equation}
    \Delta\nu = 2\sqrt{\ln{2}}\frac{D\nu}{B_{\rm max}}\beta_{\rm max},
\end{equation}
where $D$ and $B_{\rm max}$ represent the antenna diameter and maximum interferometer baseline lengths respectively, $\nu$ is the observing frequency, $\Delta\nu$ is the maximum beam smearing permitted for the value $\beta_{\rm max}$, determined by the fractional loss in peak emission intensity at the half-power-beam-width ($R_{\Delta\nu}$) for which
\begin{equation}
    R_{\Delta\nu} = \frac{I}{I_0} = \frac{\sqrt{\pi}}{2\sqrt{\ln{2}}\beta_{\rm max}} \rm{erf}\big( \sqrt{\ln{2}}\beta_{\rm max}{} \big),
\end{equation}
(we note that the units of $\nu$ and $\Delta\nu$, and also $D$ and  $B_{\rm max}$ must match).
For a relatively low fractional loss of ${<}5$\%, one can show that $\beta_{\max}{\approx}0.48$, and thus these SMA observations with 6\,m antennas, projected VEX uv-baselines of 508.9\,m, and a mean frequency of 225.538\,GHz allow for channel averaging up to this beam smearing criterion of ${\approx}2\,$GHz. 
The averaging we applied provides channels with widths of 0.5\,GHz, and thus a factor of 4 smaller than this limit, and thus well inside acceptable ranges.
Inverting our channel averaging choice, one can show that the choice of channel averaging will lead to bandwidth smearing at the level of ${\approx}0.3\%$. 
We combined these three measurement sets with the {\tt CASA concat} package. 
Self-calibration was attempted on the combined measurement set, however this failed to improve the image fidelity/noise properties (likely due to the relatively low SNR versus that needed for successful self-calibration). 

\begin{figure}
    \centering
    \includegraphics[width=0.475\textwidth]{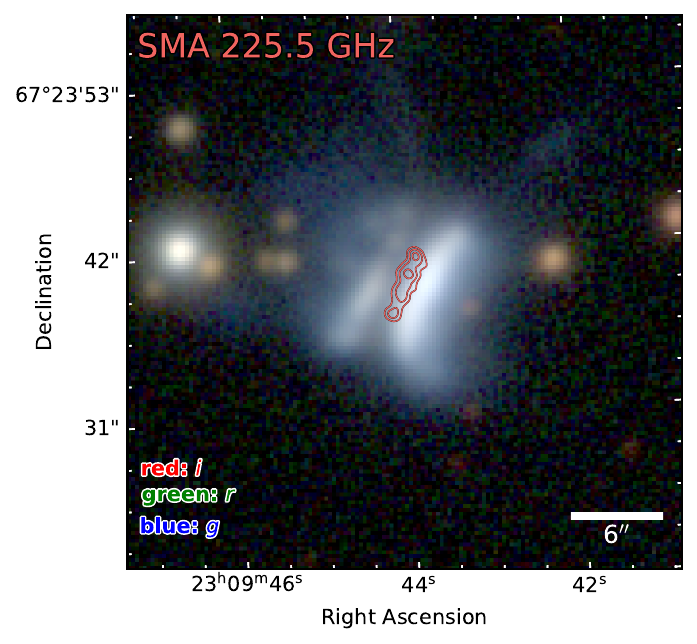}
    \caption{Pan--STARRS RGB color map \citep[as presented in][]{Monsch24} with the newest SMA high-resolution continuum contours over-plotted (at the 8, 12 and 16$\sigma$ levels). }
    \label{fig:RGB}
\end{figure}

Combined, the three SMA configurations provide projected uv-baselines spanning 5.8--508.9\,m, capable of a maximum angular resolution of ${\sim}0\farcs5$ (along the beam minor axis), whilst providing very good sensitivity to emission on angular scales of ${\sim}23''$ \citep[i.e., for a 50$^\mathrm{th}$-percentile flux reconstruction, see eq.~A11 and eq.~1 of][respectively]{Wilner1994, Lovell25}.
We present in the left panel of Fig.~\ref{fig:main} the SMA continuum image with a sensitivity of ${\sim}220\,\mu{\rm Jy\,beam}^{-1}$, the same contours of this image over-plotted on the RGB map as presented by \citet{Monsch24} in Fig.~\ref{fig:RGB}, and describe in \S\ref{sec:analysis} the procedure adopted to image the 1.3\,mm data.

\subsection{NOEMA observations \& data calibration}
\label{sec: NOEMA data}
We present new NOEMA 2.7--3.1\,mm observations of IRAS~23077+6707 (project W23BJ, PI: L. Trapman). These observations were carried out between 27 December 2023 and 4 March 2024 in configurations C (24--368\,m; one track) and A (72--1768\,m; ten tracks).
These configurations are capable of a maximum angular resolution of ${\sim}0\farcs34$ (along the beam minor axis), and provide very good sensitivity to emission on angular scales of ${\sim}25''$ (also for a 50$^\mathrm{th}$-percentile flux reconstruction).
In aggregate the total time on source was 17.6\,hrs, resulting in typical sensitivity of ${\sim}11{-}15\,\mu{\rm Jy\,beam}^{-1}$.
Additional observing details are presented in Appendix~\ref{app:B}, Table~\ref{tab:obs}.

The \texttt{POLYFIX} correlator was set up with low-resolution (2\,MHz) spectral windows covering 92.19--100.316 GHz in the lower side band (LSB) and 107.678--115.804 GHz in the upper side band (USB). 
Additional high-resolution (62.5\,kHz) spectral windows were included covering various lines, including CO $J{=}1{-}0$ isotopologues. 
Except for the C$^{18}$O $J{=}1{-}0$ line (which we discuss further in \S\ref{sec:analysis}), we only analyse the low-resolution spectral windows in this work.

The data were calibrated using the standard NOEMA pipeline in the \texttt{CLIC} package. 
After calibration, all low spectral resolution chunks of calibrated science data were collated into two uv-tables for the LSB and USB data respectively centered at frequencies of 96.254\,GHz (3.1\,mm) and 111.74\,GHz (2.7\,mm). 
We masked spectral chunks within 30 km/s of CO isotopologue lines and spurious lines detected during pipeline calibration of the data. 
After masking, the LSB and USB datasets were averaged along their spectral axes. 
We present in the central and right panels of Fig.~\ref{fig:main} the USB and LSB NOEMA continuum images with sensitivities of ${\sim}15\,\mu{\rm Jy\,beam}^{-1}$ and ${\sim}11\,\mu{\rm Jy\,beam}^{-1}$ respectively, and describe in \S\ref{sec:analysis} the procedure adopted to image the 2.7\,mm and 3.1\,mm data.

\section{Image analysis} \label{sec:analysis}
\subsection{The bulk disk}
We present in Fig.~\ref{fig:main} a 3-panel gallery of IRAS~23077+6707, showing high-resolution SMA and NOEMA (USB and LSB) maps, all with beam major axes of $0\farcs8$\footnote{We host these three final calibrated continuum images on Zenodo \citep{LovellData_Dracula2025} to enable others to utilize these images for analysis.}.
We selected ${\sim}0\farcs8$ to present these images, as we found this provided the best trade-off between resolution and signal-to-noise for the three images, whilst keeping these at the same effective resolution.
All images were cleaned to a threshold of $1{-}2\sigma$ based on the theoretical noise values as measured by {\tt CASA} and {\tt GILDAS} during cleaning, and with comparable resolutions (with beam major and minor axes of $0\farcs80{\times}0\farcs69$, $0\farcs80{\times}0\farcs76$ and $0\farcs81{\times}0\farcs77$ respectively, and beam position angles of $61^\circ$, $50^\circ$, and $60^\circ$ respectively).
The SMA, NOEMA USB and LSB images achieve RMS sensitivities of $\sigma_{\rm SMA}=220\,\mu$Jy\,beam$^{-1}$, $\sigma_{\rm USB}=15\,\mu$Jy\,beam$^{-1}$ and $\sigma_{\rm LSB}=11\,\mu$Jy\,beam$^{-1}$ respectively.
The SMA data required a `briggs' weighting scheme with robust parameter 0.75, and the NOEMA data required a GILDAS-based robust weighting of 5.6 and 10 (with uv-tapers of 880m${\times}$550m, 10\,deg and 560m${\times}$520m, 45\,deg for the LSB and USB, respectively). 
These images are the first to provide sub-arcsecond resolution of IRAS~23077+6707, constituting a five-fold increase in the resolution of published 1.3\,mm data \citep{Monsch24}, and are the first presented observations at 2.7--3.1\,mm.

In all three cases, these images resolve the disk midplane radially, which has the typical features of highly-inclined protoplanetary disks at millimeter wavelengths, whereby the cold mid-plane emission is seen as a linear `cigar'-like structure along the disk major axis \citep[see e.g.,][]{Villenave2020}.
From $4\sigma$ contours of these images, we measure emission extents ranging from $5\farcs6{-}6\farcs1$.
These angular scales are ${\sim}5{\times}$ smaller than the 50$^\mathrm{th}$-percentile flux reconstruction (provided in \S\ref{sec:data}) and thus we expect this size range reflects the true angular extent of the millimeter emission (i.e., the data are not resolving out larger-scale emission).

We derive basic disk properties by fitting 2D Gaussian ellipses to these three images in {\tt carta} \citep{carta} within a rectangular region centred on the image peak with a length and width of $10''$ and $6''$ respectively. 
We present the results in Table~\ref{tab:2dfit}.
All three image residuals demonstrate that a simple Gaussian ellipse is not a good structural match to the disk emission.
Nevertheless, 2D Gaussian fits still provide reasonable constraints on the total integrated disk flux, position angle, and (from the resolved major and minor FWHM measurements) a crude lower-limit estimate of the disk inclination, which we find ranges between 68--73$^\circ$.
We verified this approach by confirming the angle of the disk spine agreed with each fit to the PA, and by measuring the total flux within a $10''{\times}6''$ box centered on the disk within each image; in all three cases these agreed within the stated errors.
From these major and minor axis FWHM values, we derive geometric-based lower-limits on the disk inclination based on $i=\arccos[{\rm{FWHM_{minor}/FWHM_{major}}}]$ for each of the three 2D fits. 
We note these are only lower-limits given that such a geometric estimate neglects any vertical disk extent (which would enhance the minor axis extent).

\begin{table}[]
    \centering
    \caption{Best-fit 2D Gaussian parameters to the three images of IRAS~23077+6707. Formally, the fitted errors for the three images were 2.4\,mJy, 0.12\,mJy and 0.10\,mJy (SMA, USB and LSB respectively), however we report here the flux uncertainties including flux calibration uncertainties of 10\% added in quadrature (for NOEMA; see \url{https://www.iram.fr/IRAMFR/GILDAS/doc/html/pdbi-cookbook-html/node17.html} for details for 3\,mm data). }
    \begin{tabular}{c|ccccc}
      \hline
      \hline
      Image & Flux & Major axis & Minor axis & PA  \\
       & [mJy] & [$''$] & [$''$] & [$^\circ$] \\
      \hline
       SMA & $49.3{\pm}5.5$ & $4.62{\pm}0.22$ & $1.70{\pm}0.07$ & $335.5{\pm}1.4$ \\
       USB & $3.17{\pm}0.34$ & $4.58{\pm}0.18$ & $1.38{\pm}0.04$ & $334.6{\pm}0.8$ \\
       LSB & $2.01{\pm}0.22$ & $4.54{\pm}0.22$ & $1.56{\pm}0.06$ & $336.8{\pm}1.3$ \\
      \hline
      \hline
    \end{tabular}
    \label{tab:2dfit}
\end{table}

Fig.~\ref{fig:main} shows that in all cases the phase center of the observations (\textit{0,0} offset coordinates) does not align with either the geometric center of the disk, nor the emission peak.
This apparent difference is likely the result of the central star being occulted by the disk.
Since these coordinates are not well-centered on the disk, we utilize the relatively symmetric C$^{18}$O $J{=}1{-}0$ line data to derive a `dynamic' disk center, with which we conduct analysis with later on. 
We define this dynamic center as the intersection of the peak continuum disk spine and the iso-velocity contour at $1.8$\,km\,s$^{-1}$ (i.e., the systemic velocity of IRAS~23077+6707; see further details in Appendix~\ref{app:A}).
We thus define a dynamic center shifted eastwards (x) and northwards (y) from the imaged phase center by $+0\farcs18$ and $+0\farcs52$ respectively.

With three fluxes (and images) at different wavelengths we can measure the spectral slope from 1.3--2.7\,mm and 1.3--3.1\,mm.
From the integrated fluxes, we measure steep, mean spectral slopes of $\alpha{=}d\log{F_\nu}/d\log{\nu}$, of $\alpha_{1.3{-}2.7}{=}3.92{\pm}0.22$ and $\alpha_{1.3{-}3.1}{=}3.75{\pm}0.19$.
We find by producing a clipped mask (based on regions where all three images have emission exceeding $5\sigma$) little radial variation in the spectral slope across the resolved maps of $\alpha$ through the disk, with a lower mean $\alpha$ by 0.3 in both 1.3--2.7\,mm and 1.3--3.1\,mm maps.
It is plausible that the value of $\alpha$ is being skewed to large values by the 1.3\,mm SMA flux. 
Archival SMA data in the SMA compact configuration exist for IRAS~23077+6707 at 264\,GHz (1.1\,mm) and 334\,GHz (0.90\,mm), with associated fluxes of $51{\pm}5$\,mJy and $115{\pm}9$\,mJy respectively (project 2024A-S002, PI: J. B. Lovell).
Whilst we do not present these images here due to their lower angular resolution, with these fluxes we instead measure a range of $\alpha$ values from 3.2--3.5, with associated errors in the range 0.1--0.7, i.e., $\alpha_{0.9{-}2.7}{=}3.28{\pm}0.14$, $\alpha_{0.9{-}3.1}{=}3.25{\pm}0.13$, $\alpha_{1.1{-}2.7}{=}3.23{\pm}0.18$, $\alpha_{1.1{-}3.1}{=}3.21{\pm}0.16$, and $\alpha_{0.9{-}1.1}{=}3.5{\pm}0.7$.
This suggests that the $\alpha$ derived with the SMA 1.3\,mm data is discordant at the ${\sim}1\sigma$ level, and perhaps slightly overestimating the spectral slope.
As such we report the complete range of derived $\alpha$ consistent with these data to span from 3.2--3.9. 

\begin{figure}
    \centering
    \includegraphics[clip, trim={0cm 0.0cm 0cm 0.975cm}, width=1.0\linewidth]{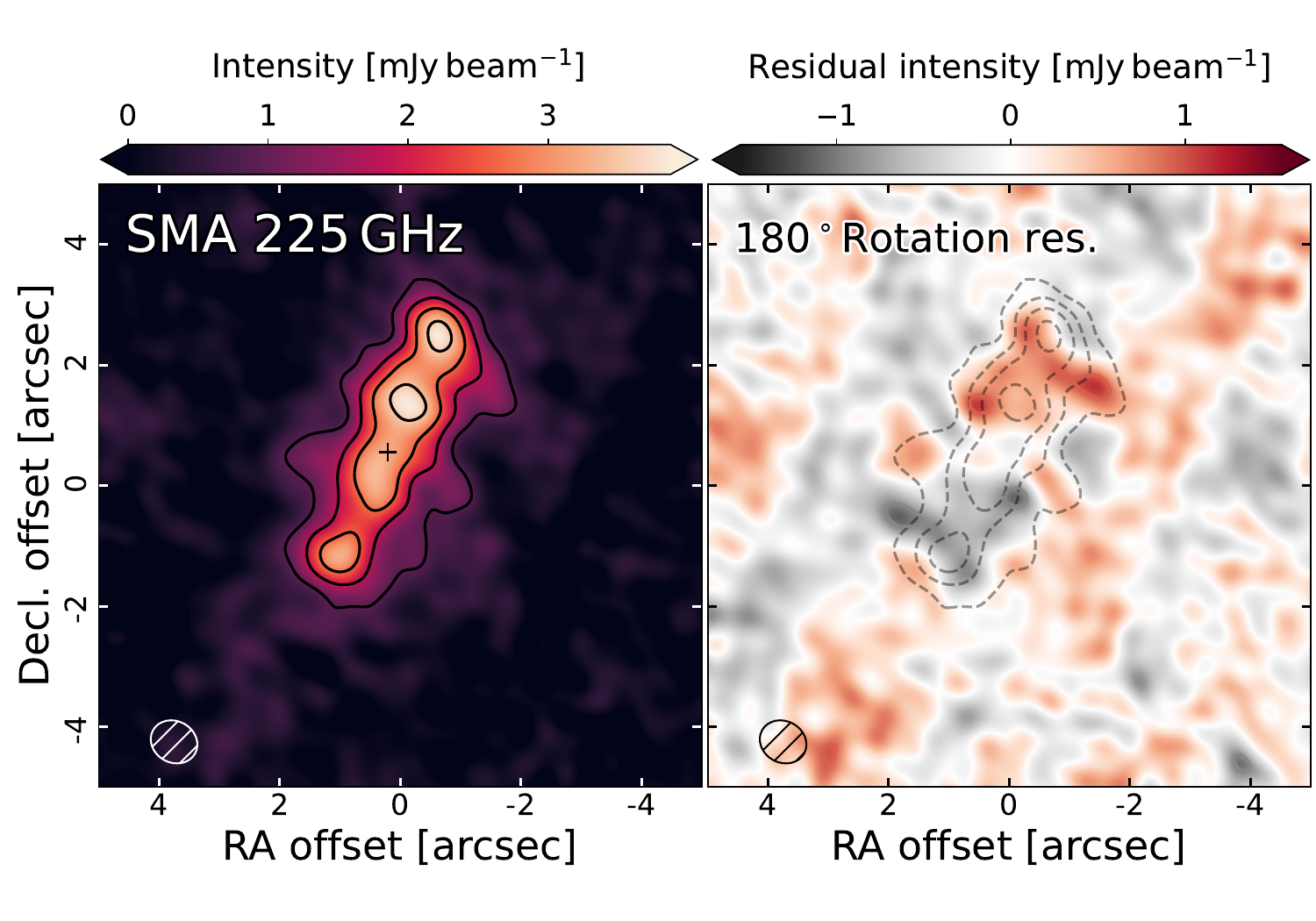}
    \includegraphics[clip, trim={0cm 0.0cm 0cm 0.975cm}, width=1.0\linewidth]{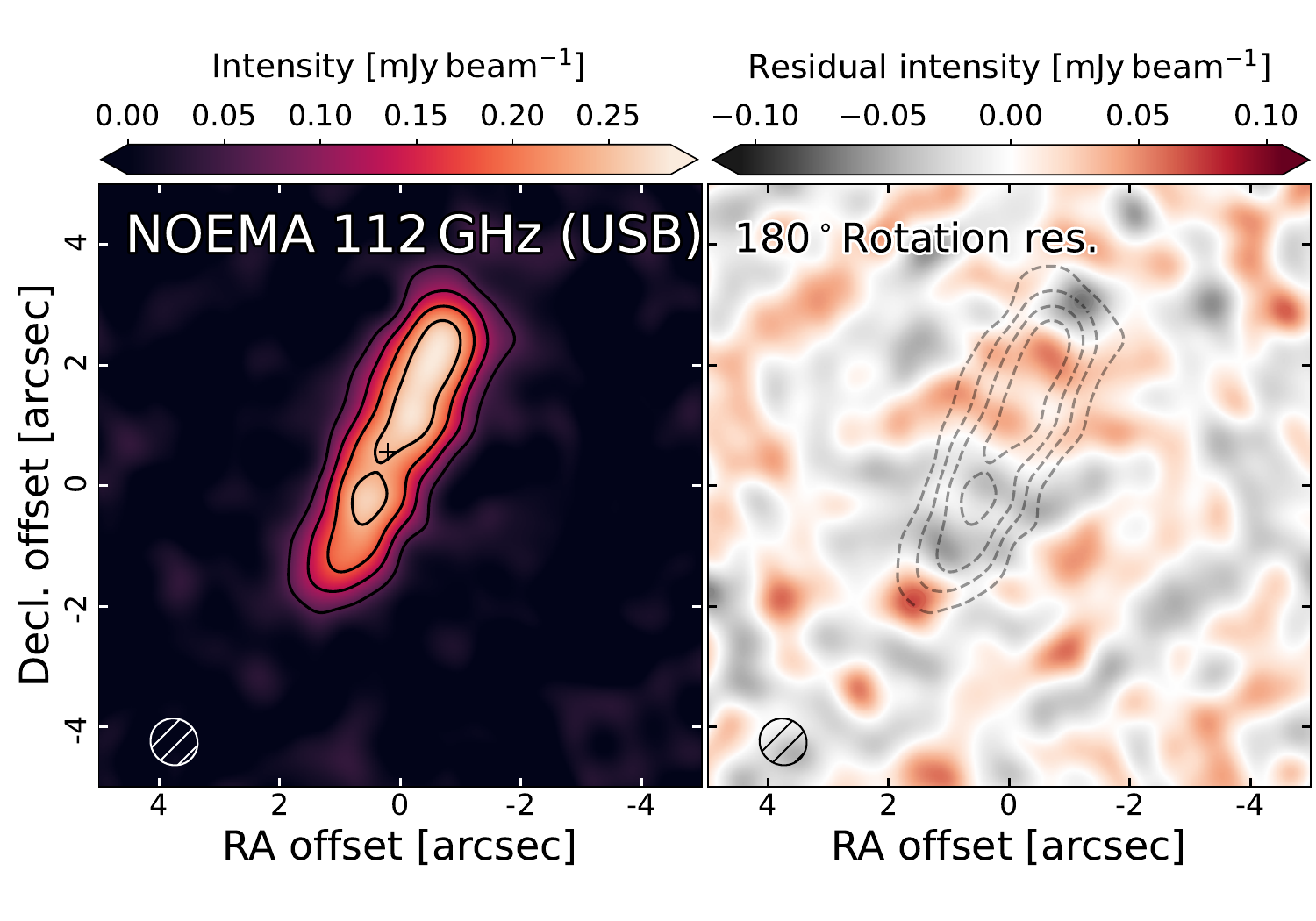}
    \includegraphics[clip, trim={0cm 0.0cm 0cm 0.975cm}, width=1.0\linewidth]{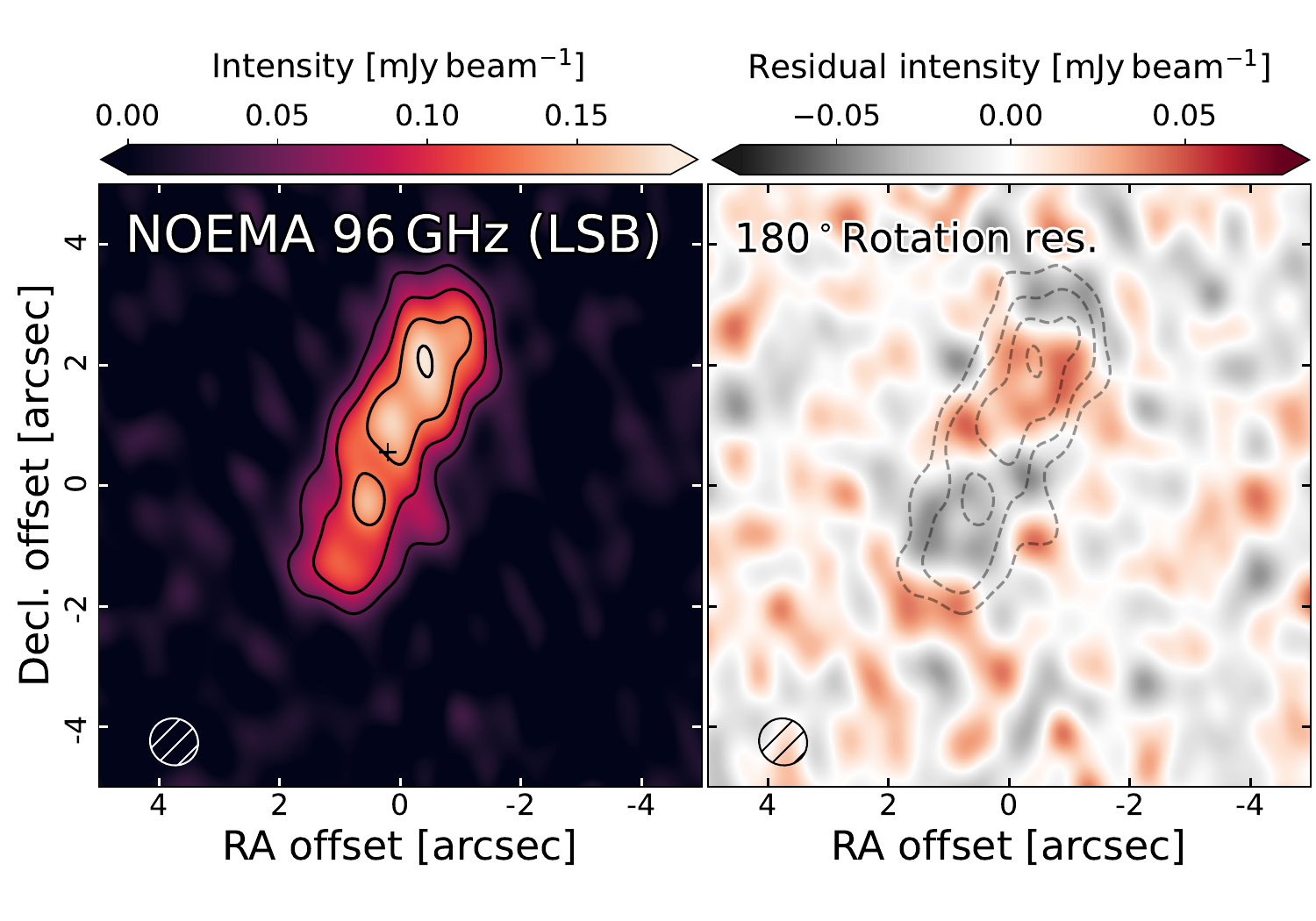}
    \caption{Self-subtraction maps for the SMA (top), NOEMA USB (middle) and NOEMA LSB (bottom) images. We present in the left panel each image and in the right panel each residual map, which we produce by subtracting from each image, the same image but $180^\circ$-rotated about the dynamic center. Contours in the image maps match those of Fig.~\ref{fig:main} which are imprinted on the residual maps with gray-dashed lines. Residual maps show ${\pm}3, {\pm}5\sigma$ residual image rms contours (formally, these are $\sqrt2\times$ each image's rms).}
    \label{fig:rotSubs}
\end{figure}

\subsection{Radial, asymmetric emission}
\label{sec:asymmetric_emission}
Having characterised basic properties of the bulk disk emission, here we quantify the asymmetric features in the data.
We note two features evident in all three images, a north--south intensity enhancement, with brighter emission always appearing in the north, and a north--south total flux asymmetry, favoring the north.
To derive north--south total flux ratios, we measure the total flux in the north and south of the images within the top and bottom halves of a $10{\times}5''$ rectangular aperture centered on the disk dynamic center, rotated to a position angle of $335.5^\circ$.
We find north--south total flux ratios for the SMA, NOEMA USB and NOEMA LSB of ${\sim}1.14$, ${\sim}1.32$ and ${\sim}1.35$ respectively, with estimated errors at the few-percent level.
These derived values suggest that the longer-wavelength data is more sensitive to the total flux asymmetry (where this is significantly stronger), which may be due to the better signal-to-noise in the NOEMA data, or the lower continuum optical depth from 2.7\,mm--3.1\,mm emission versus that at 1.33\,mm.

Following a similar method adopted by \citet{Lovell21c} where they investigated emission asymmetries in the debris disk of HD\,10647 (q$^1$\,Eri), we produce a series of image-plane self-subtraction maps of the data to understand the significance of the brightness asymmetries.
We show in Fig.~\ref{fig:rotSubs} self-subtracted image residuals, obtained by rotating each image by $180^\circ$ (about the dynamic center\footnote{If we instead conduct this exercise about the phase center, the residual emission maps show much stronger residual emission structures, which typically indicates that there is a geometric offset between the rotation center and the true disk center.}) and subtracting this rotated image from the original image (a process that results in residual image rms errors of $\sqrt{2}$ larger than the data). 
Each residual map clearly demonstrates that significant intensity differences are present between the north and south, by hosting highly significant residual contours, and the presence of broad positive/negative regions that span multiple independent beams over the region of the disk. 
Although the residual peaks in the 1.3\,mm SMA data are the largest in absolute terms, these appear to be consistent to the NOEMA residuals in relative terms, despite there being an order of magnitude difference in the mean disk intensity/total flux between the 1.3\,mm and 3.1\,mm data.
We quantify this by defining a flux-scaled asymmetry parameter, `$A$'
\begin{equation}
    A_{\rm \{ dataset \}} = \mathcal{M}_{i,j}\frac{\sum_{i,j}|I^{\rm resid}_{i,j}|}{\sum_{i,j} I^{\rm image}_{i,j}}
\end{equation}
which is simply the ratio of the absolute total flux in the residual map  ($\sum_{i,j}|I^{\rm resid}_{i,j}|$) to the total flux in the image ($\sum_{i,j} I^{\rm image}_{i,j}$) within a masked region ($\mathcal{M}_{i,j}$) which we define based on all regions of the image that exceed $5\sigma$ (removing any intrinsically noisy data from our calculation).
For the SMA, NOEMA USB and LSB data we measure $A_{\rm SMA} \approx 0.21$, $A_{\rm USB} \approx 0.14$ and $A_{\rm LSB} \approx 0.22$, which suggests the relative levels of non-axisymmetric emission are moderate in comparison to the bulk disk emission\footnote{We investigated different masking thresholds from 4--6$\sigma$ and found $A_{\rm \{ dataset \}}$ values varied by less than 0.01.}.
We note that this type of analysis has a rich history in studying the asymmetric nature of galaxies \citep[see e.g.,][]{Conselice+2000,Conselice2003, Davis+2022} where the asymmetry parameter has been referred to as the `concentration-asymmetry-smoothness parameter' ($A_{\rm CAS}$).

In comparison to the lower-resolution continuum maps presented in \citet{Monsch24} which spanned angular scales comparable to the Pan--STARRS scattered light data, we show in Fig.~\ref{fig:RGB} that the resolved continuum emission is approximately half the angular extent of the scattered light.
These latest resolved maps thus agree with the \citet{Monsch24} 2--3$\sigma$ radii estimated based on simple image de-convolution analysis.
Perhaps more interesting to compare however are the scattered light asymmetries and millimeter continuum which appear anti-aligned.
For example, whilst the northern region of millimeter disk is brighter than the southern region, \citet{Monsch24} state in relation to the optical data that ``\textit{the eastern (fainter) lobe shows a north–south asymmetry, with the southern region being brighter by a factor of 3}''.
Whilst more subtle, there is further evidence that in the optical data the western (brighter) lobe also has a north–south asymmetry, favoring instead the north.
\citet{Monsch24} discuss the possibility that these features are due to a misaligned inner disk, which we consider further in \S\ref{sec: origin of the asymmetries} given the possibility this scenario can also influence the morphology of the millimeter emission.

\begin{figure*}
    \centering
    \includegraphics[clip,trim={0cm 0.2cm 0cm 0.1cm}, width=1.0\linewidth]{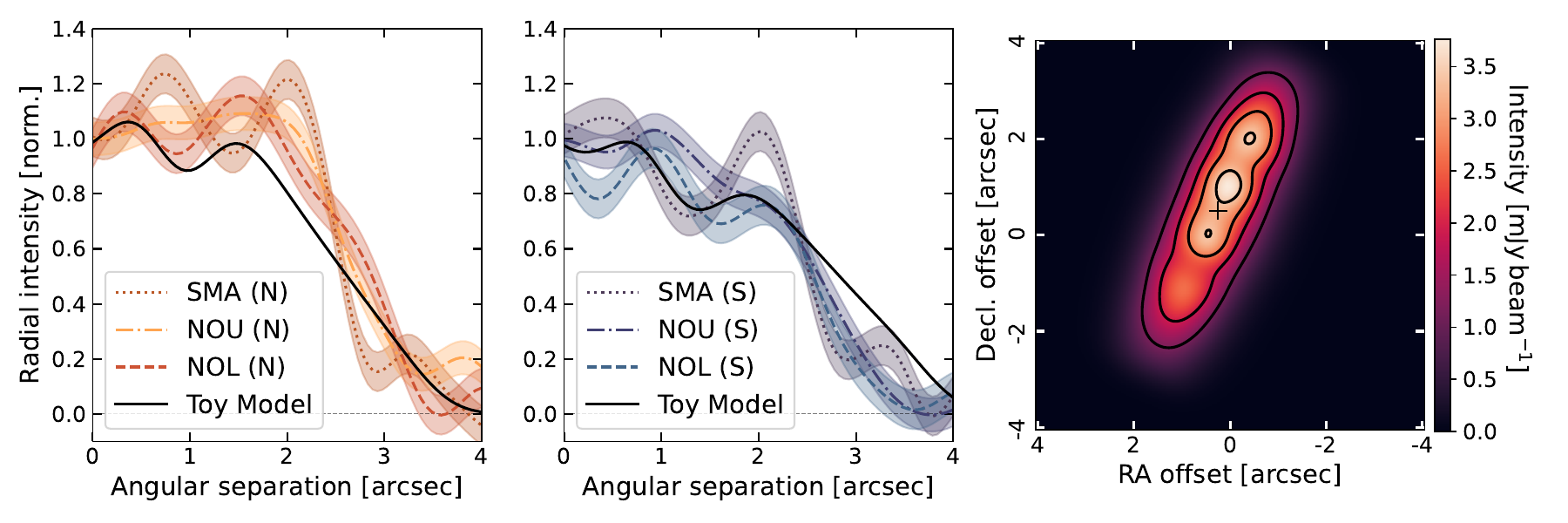}
    \caption{Left and center: Radial intensity profiles of the SMA and NOEMA data (where NOU is for the NOEMA USB, and NOL is for the NOEMA LSB) for the northern (N) and southern (S) halves of the disk. The toy model profile is over-plotted on these radial profile plots with a black line. Shaded regions represent the $1\sigma$ RMS uncertainty on the profile. Right: toy eccentric model map (simulated at 1.3\,mm) with contours shown at 30\%, 50\%, 70\% and 90\% of the model peak.}
    \label{fig:radProfs}
\end{figure*}

\newpage
\section{Discussion} \label{sec:discussion}
\subsection{Comparison to the resolved millimeter population of highly-inclined/edge-on disks}
We have derived a range of bulk disk quantities such as the flux, radial extent and spectral index ($\alpha$) that we discuss here in comparison to the (millimeter wavelength) resolved population of protoplanetary disks.
Firstly, considering the angular scale of all protoplanetary disks in either the `DSHARP' sample \citep{Andrews18} and the full population of disks in the Protostars and Planets chapter~15 \citep{Manara23}, IRAS~23077+6707's continuum extent presents significantly larger.
In comparison to the ALMA-studied sample of 12 edge-on disks in \citet{Villenave2020}, IRAS~23077+6707 is more radially extended along its major axis than all sources, except for IRAS~04158+2805.
Whilst higher resolution (sub)-millimeter data for IRAS~18059-3211 (GoHam) is not publicly available, we note that the lower-resolution SMA continuum images presented in \citet{Teague2020} show the source to extend $\approx5{-}6''$, thus comparable with IRAS~23077+6707.
As such, IRAS~23077+6707, along with IRAS~18059-3211 and IRAS~04158+2805 appear to be extremes in the population of protoplanetary disks, with extensions all in excess of $5''$.
That the three largest extent protoplanetary disks are all highly-inclined/edge-on is unlikely a coincidence, given their optical depths and geometries, it is plausible that edge-on disks are more easily detected out to larger radii than face-on disks.

IRAS~23077+6707 has a steep spectral index (we derive values between 3.2--3.9 based on different total flux measurements), in excess of the bulk population for protoplanetary disks, e.g., compared with $\alpha=2.2{\pm}0.2$ (for $\lambda=0.9{-}1.3\,$mm) or $\alpha\approx2{-}3$ (for $\lambda=1{-}3\,$mm), see e.g., \citet[][]{Andrews20}, \citet{Tazzari21} and \citet{Chung+2024}. 
The values we derive for IRAS~23077+6707 are also at the upper-end in comparison to the sample of highly-inclined/edge-on protoplanetary disks \citep[which range from 2.1--2.9, see][]{Villenave2020}.
Whilst the values of $\alpha$ are not unprecedented, reconciling the broad range of $\alpha$ with new data and/or simulations is an important next step to understand if this steep spectral index is due to limitations in the SNR of the data, or intrinsically different physical properties, e.g., a large population of small grains resulting from a lack of grain growth, and thus how this compares to the broader disk population.

Whilst IRAS~23077+6707 stands out in terms of its spectral slope and disk extent, the source is rather nominal in terms of its total integrated flux (which we note remains true for any of the SMA or NOEMA flux measurements).
We compare the measured SMA 1.3\,mm flux with the luminosity distribution presented in \citet{Andrews20}, scaling the flux to an estimated 0.9\,mm luminosity density ($L_{\rm 0.9\,mm}$) by extrapolating the spectral slope.
If IRAS~23077+6707 is located at a distance of 150\,pc, then its $L_{\rm 0.9\,mm}$ falls almost perfectly on the expected $L_{\rm 0.9\,mm}$--$M_\star$ distribution for 1--2\,$M_\odot$ systems. 
IRAS~23077+6707 is located in the Cepheus star-forming region, with a range of estimated distances to the nearest sub-regions from 150--370\,pc \citep{Szilagyi21}.
If instead, IRAS~23077+6707 is located towards these furthest distances, then its millimeter luminosity would be ${\approx}6{\times}$ higher, and present in the uppermost location of the $L_{\rm 0.9\,mm}$--$M_\star$ distribution.

Finally, unlike the vast majority of highly-inclined/edge-on disks, the images that we present of IRAS~23077+6707 all show evidence of a large-scale radial asymmetry.
The asymmetry present in the disk of IRAS~23077+6707 is prominent as a brightness enhancement in the north of the disk, evident in all three images in Fig.~\ref{fig:main} and  the self-subtracted maps in Fig.~\ref{fig:rotSubs}, and as a total flux asymmetry (favoring the north).
We note in the case of \citet{Villenave2020}, just one source hosted a significantly asymmetric radial profile; the disk of IRAS~04158+2805.
Just as in these data for IRAS~23077+6707, the source IRAS~04158+2805 was first shown by the SMA to host a strong radial asymmetry in \citet{Andrews+08}.
Subsequently that system was resolved by ALMA as an eccentric circumbinary ring \citep[see][]{Ragusa+2021}, which offers one plausible interpretation for the disk of IRAS~23077+6707.
\citet{Teague2020} likewise note the presence of a north--south continuum asymmetry in the continuum data for IRAS~18059-3211, further suggesting an interesting concordance of strong brightness asymmetries in this small population of large, highly-inclined disks.

\subsection{IRAS~23077+6707's disk asymmetric origins}
\label{sec: origin of the asymmetries}
In \S\ref{sec:analysis} we presented evidence of strong brightness and total flux asymmetries in IRAS~23077+6707's disk (favoring a brighter disk in the north), and highlighted how these compare to the scattered light asymmetries presented by \citet{Monsch24}.
Here we discuss a number of plausible scenarios that may produce such emission asymmetries, none of which we can yet rule out, including the possibility that IRAS~23077+6707 hosts an eccentric disk, among other hypotheses.

\subsubsection{An eccentric disk?} \label{sec:eccdisk?}
We have noted a number of strong similarities between IRAS~23077+6707 and IRAS~04158+2805 (their sizes, spectral slopes, and radial asymmetries) which we provide as a basis to suggest one plausible interpretation that IRAS~23077+6707 hosts an eccentric disk.
To investigate the hypothesis that IRAS~23077+6707 is host to an eccentric disk, we construct a simple, toy model in {\tt RADMC-3D} \citep{Dullemond12}, utilizing the models of \citet{LynchLovell21} and \citet{LovellLynch2023}.
These models make a number of assumptions given these were designed to model eccentric debris disk dust, e.g., these explicitly assume the dust continuum is optically thin, and that the dust dynamics are independent of any gas.
To model the system, we define a parametric surface density function that radially comprises of two top-hats that are distributed azimuthally around eccentric orbits (with a negative gradient eccentric power-law $e{\sim}a^{-1}$, and pericenter direction along the disk spine south-wards with $\omega_f=0.0^\circ$), around a hot ($T=8000\,$K) star \citep[consistent with the analysis IRAS~23077+6707's optical spectrum in][]{Berghea24}. 
The separation and widths of the top-hats ensure that the intensity peaks are radially resolved (as in the data) with a local minima near to the center of the disk spine.
We remain agnostic as to the distance to IRAS~23077+6707 (and thus absolute disk radii and mass), and simulate the scenario at an arbitrary distance of 20\,pc, fixing the radii and widths of the top-hats to match the peaks in the SMA and NOEMA data, the inclination to $75^\circ$, the position angle to $335.5^\circ$, the (Gaussian) scale height to $H=0.1{\times}r$, and the offset to the coordinates of the dynamic center.

We simulate this scenario at 1.3\,mm (matching the SMA data) with {\tt RADMC-3D} \citep{Dullemond12}, and present both the 2D image in the right panel of Fig.~\ref{fig:radProfs}, and normalised radial profiles of this model in the left and center panels, along with those of the three images\footnote{Radial profiles were constructed along the disk spine (parallel to the derived position angles) relative to the dynamic center, and averaged over ${\pm}0.3''$ along the minor axis. 
We note that peaks in the SMA and NOEMA (LSB) profiles appear out of phase. We attribute this apparent phase difference to interferometric imaging artifacts which partially result from the tapering of the NOEMA LSB data (in the natural-weighted LSB image, these profile peaks line up more consistently), as well as owing to the uv-samping of the SMA and NOEMA observations which yield beams aligned almost parallel to the disk minor axis, but with different widths.}.
Physically, this setup describes two eccentric rings, with a cavity in the central region near to the location of the star/stars.
Whilst the model is not a perfect fit to the data, despite the simplistic assumptions we make, this model nevertheless shows remarkably good consistency with the 2D distribution of emission including the intensity structure along the disk spine for the SMA and NOEMA data, and the total 1.3\,mm flux.
The model we present has an eccentricity of $e=0.26$, moderately high in comparison to that of the protoplanetary disk HD~100546 \citep[see][who derive a value of $e\approx0.07$ for HD~100546's outer ring]{Fedele+2021}, but consistent with values derived for other protoplanetary disks \citep[see e.g.,][in the case of IRS~48, $e\approx0.27$]{Yang+2023} and debris disks \citep[see e.g.,][in the case of HD~53143, $e\approx0.21$]{Macgregor22, LovellLynch2023}. 
By comparing to models with zero eccentricity, we found that this level of eccentricity was essential to produce the north-south asymmetric intensity ratios along the disk spine.
We avoid fitting this model to derive best-fit parameter estimations (given the data quality, there are likely many degenerate solutions), but infer that if the disk is eccentric, its eccentricity needs to be in the vicinity of $e\approx0.2{-}0.3$ to reproduce the observed millimeter flux asymmetry.

We do not simulate the scenario at optical wavelengths, however note that the expected enhancement in mid-plane dust density at disk ansae in an eccentric disk scenario (for disks with apocenter directions aligned towards the north) could plausibly obscure more scattered light preferentially in the north of the disk (where dust at apocenter would be more vertically raised versus pericenter), and induce such an anti-aligned scattered light asymmetry discussed in \S\ref{sec:asymmetric_emission}.
Nevertheless, given that optical and millimeter wavelengths trace altogether different regions of protoplanetary disks, such an anti-alignment may equally have resulted from entirely different physical processes.
Indeed, comparisons to studies of eccentric protoplanetary disks shows there is no clear correspondence between scattered light emission/asymmetries and those observed at millimeter wavelengths.
For example, in the case of HD~100546, \citet{Fedele+2021} present SPHERE images overlaid with ALMA $870\,\mu$m contours and show the eccentric (sub-mm) ring emission that peaks in the north has no corresponding J-band emission peak, and is host to a bright spiral arm in the south of the disk, which likely resides in a much higher vertical layer of the disk.
In the case of the eccentric disk Oph~IRS~48, the combined studies of \citet{Yang+2023} and \citet{vdMarel+13} show this system to be eccentric, as well as being host to a bright southern arc at millimeter wavelengths, that has no corresponding 18.7$\mu$m (VLT VISIR) enhancement in the mid-infrared at the same disk location. 
In the more inclined disk of IRAS~04158+2805, although the scattered light morphology has a strong north-south (geometric) asymmetry favoring the northern direction \citep{Andrews+08}, the sub-millimeter ALMA images show an east-west asymmetry in the disk \citep{Ragusa+2021}.
Overall therefore, even in well-known eccentric protoplanetary disks it appears relatively common to find little-to-no apparent correspondence between emission structures traced by micron-sized grains and larger millimeter-sized grains, the anti-alignment observed here appears neither unusual, nor definitively suggestive that the observed asymmetries need be physically connected.

\subsubsection{Alternative scenarios?}
In principle, there are many other known disk sub-structures that can induce brightness/total flux variations radially, especially in edge-on scenarios (where regions of the disk midplane may be obscured from view).
Some such scenarios observed in protoplanetary disks such as arcs \citep[see e.g.,][suggestive of the presence of hydrodynamic vortices]{Dong+2018, Cazzoletti+2018, Perez+2018, Isella+2018}, and spirals \citep[see e.g.,][suggestive of the presence of gravitational instabilities]{Benisty+15, Benisty+2017, Huang+18b}, and in earlier stage systems, large-scale offsets between the disk center along the major axis \citep[see e.g.,][that may also result from some form of gravitational instabilities]{Xu+2023, Kido+2023, Takakuwa+2024}.
Other alternatives include the possibility that there exists a misaligned inner disk, which simulations have shown can induce strong millimeter asymmetries in disks \citep[see e.g.][]{Facchini+2018}, as already suggested as one interpretation of IRAS~23077+6707's disk by \citet{Monsch24}, or further via dust filtration and protoplanet accretion \citep[suggested in the case of PDS~70 by][]{Hashimoto+2015}.
In these cases, multi-wavelength tracers likewise show distinct differences in the morphologies of disks at optical/near-infrared and millimeter wavelengths, similarly as discussed in \S\ref{sec:eccdisk?}.
As yet, all such substructures could plausibly be responsible for these observations, however without higher-resolution continuum data and an analysis of the gas structure and kinematics, many of these scenarios remain indistinguishable. 
Further work is now necessary to understand the true nature and origins of IRAS~23077+6707's asymmetric morphology.

\section{Summary and Conclusions} \label{sec:conclusions}
We have presented high--resolution images of IRAS~23077+6707 (`\textit{Dracula's Chivito}') with the Submillimeter Array (SMA at 1.33\,mm/225.5\,GHz) and the Northern Extended Millimeter Array (NOEMA at 2.7\,mm/111.7\,GHz and 3.1\,mm/96.2\,GHz).
IRAS~23077+6707 is a highly-inclined and newly discovered protoplanetary disk, first reported and analysed in 2024 by \citet{Berghea24} and \citet{Monsch24}.
The data we present combines multiple configurations of SMA (compact, extended, and very extended) and NOEMA (A and C), and thus images at high angular resolution, which we present at ${\lesssim}0\farcs8$. 
These images constitute the first sub--arcsecond resolution maps of IRAS~23077+6707.
We found the following: 

\begin{enumerate}[leftmargin=5.0mm]
\item The images and data show extended linear emission that spans $5\farcs6{-}6\farcs1$ radially, with a steep millimeter spectral slope ($\alpha=3.2{-}3.9$).

\item The disk spine hosts a strong asymmetry, with the northern half hosting more flux than the south, and hosting multiple brighter intensity peaks.

\item We discuss hypotheses about the potential origins of these features, including the possibility that IRAS~23077+6707 hosts a rare example of an eccentric protoplanetary disk, a misaligned inner disk, an arc or a spiral, which can induce these radially asymmetric structures. 

\item We present a very simple, toy eccentric dust continuum model of IRAS~23077+6707. We show that if the disk is eccentric, then it would need to have an eccentricity of $e\approx0.2{-}0.3$ to reproduce the bulk morphology of the emission.

\end{enumerate}

New continuum and line data to assess gas morphologies and kinematics are now needed to understand which of these hypotheses remain consistent with these continuum-based scenarios. 

\section{Software and third party data repository citations} \label{sec:cite}
\software{\tt astropy,  \citep{2013A&A...558A..33A,2018AJ....156..123A}, {\tt bettermoments} \citep{TeagueForeman-Mackey2018}, {\tt carta} \citep{carta}, {\tt CASA} \citep{McMullin+2007}, {\tt GILDAS} \url{https://www.iram.fr/IRAMFR/GILDAS}, {\tt gofish} \citep{Teague2019_gofish}, {\tt pyuvdata} \citep{Hazelton+2017, pyuvdata_Karto2025}.}
The three continuum images have all been uploaded to Zenodo \citep[][]{LovellData_Dracula2025}.

\appendix
\section{Observational setup}
\label{app:B}
IRAS~23077+6707 was observed by the SMA on three separate occasions, one per each of the SMA's Very Extended (VEX), Extended (EXT) and Compact (COM) array configurations, and on 11 occasions with two of NOEMA's configurations, once in the C configuration, ten times in the A configuration.
In Table~\ref{tab:obs} we tabulate for all fourteen observing blocks, the respective observing dates, project ID codes, integration times, calibrator sources (used to calibrate/reduce the data presented in this paper), antenna numbers (N$_{\rm{ant}}$), and the configuration. 

\begin{table*} 
    \centering
    \caption{Observational setup for the three SMA tracks and eleven NOEMA observations presented in this work. $\tau$ represents the average opacity at 225\,GHz during the SMA observations. All flux, bandpass and phase calibrators are standard SMA and NOEMA calibrator sources. Int. time represents total integrated time on source. We also specify the uv-distance ranges associated with each antenna configuration.}
    \begin{tabular}{c|c|c|c|c|c|c|c|c|c}
         \hline
         \hline
         Date & Inst. & Project & Int. & Flux & Bandpass & Phase & Antenna & N$_{\rm{ant}}$ & $\tau$ \\
         & ID & & time [hrs] & cal./s & cal. & cal./s & config. \& & &  \\
         &  & & & & & & uv-range & &  \\
         \hline
         2023 Apr 14 & SMA & 2022B-S054 & 3.8 & Ceres, & 3c279 & J0019+734, & COM & 6 & 0.049 \\
         &&&&Mars,&&J2005+778,&5.8--68.4m&& \\
         &&&&MWC349A&&J0102+584&&& \\
         \hline
         2023 Oct 07 & SMA & 2023A-S052 & 6.5 & Uranus, & BL Lac & J0102+584, & EXT & 7 & 0.100 \\
        &&&&Ceres&&J0019+734&24.6--179.2m&& \\
        \hline

         2024 Jul 07 & SMA & 2024A-S040 & 8.0 & Neptune, & 3c279 & J0102+584, & VEX & 7 & 0.112 \\  
        &&&&Ceres,&&J2005+778,&71.6--508.9m&& \\
        &&&&MWC349A&&J0014+612&&& \\

         \hline
         \hline
         2024 Dec 27 & NOEMA & W23BJ & 1.7 & LKHA101 & 3c84 & J0011+707 & C & 12 & - \\
         & & & & & & & 24--368m & & \\
         \hline
         2024 Feb 7  & NOEMA & W23BJ & 1.7 & LKHA101, & 3c273 & J0011+707, & A & 12 & - \\
                     & & &  & 2010+723 & &J0016+731 & 72--1768m& \\
         \hline
         2024 Feb 15 & NOEMA & W23BJ & 3.3 & MWC349, & 3c345 & J0011+707, & A & 12 & - \\
         & & &  & 2010+723 & &J0016+731 & 72--1768m& \\
         \hline
         2024 Feb 19 & NOEMA & W23BJ & 0.6 & LKHA101 & 3c84 & J0011+707, & A & 12 & - \\
         & & & & & &J0016+731 &72--1768m & \\
         \hline
         2024 Feb 21 & NOEMA & W23BJ & 0.34 & LKHA101 & 3c84 & J0011+707, & A & 12 & - \\
         & & & & & &J0016+731 & 72--1768m& \\
         \hline
         2024 Feb 24$^{\dagger}$ & NOEMA & W23BJ & 1.1 & MWC349, & 3c84 & J0011+707, & A & 12 & - \\
         & & &  & 2010+723 & &J0016+731 & 72--1768m& \\ 
        \hline
         2024 Feb 25 & NOEMA & W23BJ & 2.2 & MWC349, & 2200+420 & J0011+707, & A & 12 & - \\
         & & &  & 2010+723 & &J0016+731 & 72--1768m& \\ 
         \hline
         2024 Feb 26 & NOEMA & W23BJ & 2.7 & LKHA101 & 3c84 & J0011+707, & A & 12 & - \\
         & & & & & &J0016+731 & 72--1768m& \\
         \hline
         2024 Feb 27 & NOEMA & W23BJ & 1.1 & LKHA101 & 3c84 & J0011+707, & A & 12 & - \\  
         & & & & & &J0016+731 &72--1768m & \\
         \hline
         2024 Feb 28 & NOEMA & W23BJ & 1.3 & MWC349 & 3c273 & J0011+707, & A & 12 & - \\
         & & & & & &J0016+731 &72--1768m & \\
         \hline
         2024 Mar 4  & NOEMA & W23BJ & 1.5 & LKHA101 & 0923+392 & J0011+707, & A & 12 & - \\
         & & &  & 2010+723 & &J0016+731 & 72--1768m& \\
         
    \hline
    \end{tabular}
    \begin{minipage}{0.91\textwidth}
    \vspace{0.1cm}
    {\footnotesize{\textbf{Notes:} $\dagger$: An issue with the correlator setup during the observations on 24 February resulted in the exclusion of the high resolution spectral windows for this track. As data was still taken using the low spectral resolution windows used for measuring the continuum, this track is included in the data presented here.
    }}
    \end{minipage}
    \label{tab:obs}
\end{table*}

\section{Dynamical disk center via gas kinematics}
\label{app:A}

\begin{figure}
    \centering
    \includegraphics[width=0.5\textwidth]{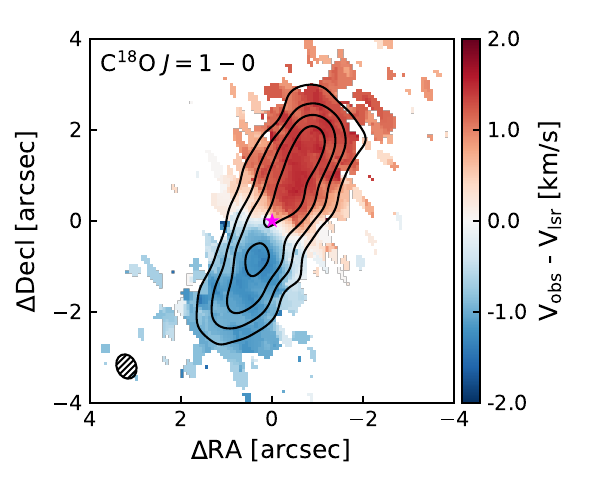}
    \caption{\label{fig: C18O moment 1 map} Moment one map of C$^{18}$O $J{=}1{-}0$. Black contours show the 112\,GHz (USB) continuum (with $4, 8, 12, 16\sigma$ contours). The magenta star denotes the dynamic center of IRAS~23077+6707 (which the map is re-centered on).}
\end{figure}

Due to its edge-on disk orientation the star (or stars), in IRAS~23077+6707 cannot be observed directly, yet its location is important for asymmetry analysis presented in \S\ref{sec:analysis}. 
The continuum asymmetry (and lack of clear stellar emission detection) renders it difficult to obtain a reliable stellar center from the continuum emission.
The C$^{18}$O line emission is far more symmetric however, which we utilize to estimate a stellar center (or barycenter). 
Figure \ref{fig: C18O moment 1 map} shows the moment one map of the C$^{18}$O $J=1-0$ emission made using {\tt bettermoments} \citep{TeagueForeman-Mackey2018}. 
Analysis of the CO isotopologue emission will be presented in a forthcoming work. 
The stellar location is determined as the intersection of the iso-velocity contour of 1.8 km/s (the velocity minima of the spectrum, which we consider as the systemic velocity of IRAS~23077+6707), and the midplane/spine of the disk taken from the continuum.
We obtained the spectrum utilizing {\tt gofish} \citep{Teague2019_gofish}, adopting a position angle of 335$^\circ$, an inclination of 75$^\circ$, and an $R_{\rm max}$ of $6''$. 
This gives a location of $(\Delta {\rm RA}, \Delta {\rm Decl}) = (+0\farcs18, +0\farcs58)$, marked as cross-hairs on Fig.~\ref{fig:rotSubs}.

\begin{acknowledgments}
\break
\textit{\large{Acknowledgments:  }}
We thank the anonymous referee for their helpful comments and suggestions on our manuscript, which greatly improved the content of this study.
We would like to thank Laure Bouscasse for help with calibrating and reducing the NOEMA data, and Pietro Curone for discussions on quantifying asymmetry parameters within images.
JBL acknowledges the Smithsonian Institute for funding via a Submillimeter Array (SMA) Fellowship, and the North American ALMA Science Center (NAASC) for funding via an ALMA Ambassadorship.
L.T. acknowledges the support of the NSF AAG grant \#2205617. 
KM was supported by NASA grants JWST-GO-1905 and JWST-GO-3523. 
The Submillimeter Array is a joint project between the Smithsonian Astrophysical Observatory and the Academia Sinica Institute of Astronomy and Astrophysics and is funded by the Smithsonian Institution and the Academia Sinica. 
The authors wish to recognize and acknowledge the very significant cultural role and reverence that the summit of Maunakea has always had within the indigenous Hawaiian community, where the Submillimeter Array (SMA) is located. 
We are most fortunate to have the opportunity to conduct observations from this mountain. We further acknowledge the operational staff and scientists involved in the collection of data presented here. 
The SMA data used here are from projects 2022B-S054, 2023A-S052, 2024A-S040. Raw (uncalibrated) data from these projects can be accessed via the Radio Telescope Data Center (RTDC) at \url{https://lweb.cfa.harvard.edu/cgi-bin/sma/smaarch.pl} after these have elapsed their proprietary access periods. 
This work is based on observations carried out under project number W23BJ with the IRAM NOEMA Interferometer. IRAM is supported by INSU/CNRS (France), MPG (Germany) and IGN (Spain).
\end{acknowledgments}

\vspace{5mm}
\facilities{Smithsonian Astrophysical Observatory (SAO)/Academia Sinica SubMillimeter Array (SMA) at Mauna Kea Observatory, NOrthern Extended Millimeter Array (NOEMA)}

\bibliography{sample631}{}
\bibliographystyle{aasjournal}
\end{document}